\documentclass[useAMS,usenatbib]{mn2e}
\usepackage{graphicx}

\title[Scenario Machine: Fast Radio Bursts, Short GRB, Dark Energy and LIGO silence]{Scenario Machine: Fast Radio Bursts, Short GRB, \\Dark Energy and LIGO silence}
\author[V. M. Lipunov, M. V. Pruzhinskaya]{V. M. Lipunov$^{1,2}$\thanks{E-mail:
lipunov2007@gmail.com}, M. V. Pruzhinskaya$^{1}$\thanks{E-mail:
pruzhinskaya@gmail.com}\\
$^{1}$Lomonosov Moscow State University, Faculty of Physics, Sternberg Astronomical Institute, Universitetsky~pr.,~13,~Moscow, 119991, Russia\\
$^{2}$``Extreme Universe Laboratory'' of Lomonosov Moscow State University, Skobeltsyn Institute of Nuclear Physics, 1(2), Leninskie gory, GSP-1, \\Moscow, 119991, Russia}

\begin{document}

\date{Accepted 2014 February 12. Received 2014 January 21; in original form 2013 December 5}

\pagerange{\pageref{firstpage}--\pageref{lastpage}} \pubyear{2002}

\maketitle

\label{firstpage}

\begin{abstract}
We discuss the recently reported discovery of fast radio bursts (FRBs) in the framework of the neutron star--neutron star (NS+NS) or neutron star--black hole (NS+BH) binary merger model. 
We concentrate on what we consider to be an issue of greatest importance: what is the NS merger rate given that the FRB rate (1/1000 yr$^{-1}$ per galaxy) is inconsistent with 
gamma-ray burst rate as discussed by Thornton and should be significantly higher. We show that there is no discrepancy between NS merger rate and observed FRB rates in the 
framework of the Scenario Machine population synthesis --- for a kick velocity of 100--150 km s$^{-1}$ an average NS merger rate is 1/500--1/2000 yr$^{-1}$ per galaxy up to z~=~0.5--1. Based on the Scenario Machine NS merger rate estimates, we discuss the lack of positive detections on the ground-based interferometers, considering the Laser Interferometer Gravitational-wave Observatory.
\end{abstract}


\section{Introduction}

The discovery of four fast radio bursts (FRBs; \cite{Thornton}) combined with the previously discussed Lorimer burst \citep{Lorimer} sparked interest in the generation mechanisms 
of extremely non-stationary and powerful radiation by neutron stars (NSs).
There are different scenarios for FRBs (see \citet{Totani} for a review) but here we discuss only NS+NS mergers, which we consider to be the most plausible explanation.

The possibility of the FRBs being generated by coalescing NSs was predicted by \citet{Lipunov96}. A sufficiently close neutron binary transforms, 
by emitting gravitational waves, into a supercompact system with a lifetime of several milliseconds (several orbital revolutions). In the process, strong electric fields are generated 
and conditions arise for acceleration of relativistic particles, which, in turn, generate non-thermal radio emission (see also \citet{Hansen, Lyutikov}).
Furthermore, \citet{Lipunova97} pointed out that a merger of NSs produces a highly magnetized and rapidly rotating object (a spinar), which can lose up to 10 per cent of its total energy 
in the form of electromagnetic radiation in the process of collapse (cf. \citet{Lipunova98, Lipunov07, Pshirkov}). However, the radiation of a 
rapidly and, possibly, differentially rotating spinar has to be highly concentrated along the axis of the object and, most likely, should be properly identified with the gamma-ray burst (GRB; \cite{Lipunov07b}). 

It is tempting to associate FRB with coalescing NSs, which, in turn, are viewed as the most likely source of short GRBs \citep{Blinnikov, Paczynski}, 
the latter, in addition, being the best source of events for gravitational-wave experiments like Laser Interferometer Gravitational-wave Observatory (LIGO), Virgo, etc. \citep{Grishchuk}. 
However, in the case of this interpretation we have to answer the fundamental question of the occurrence frequency of such bursts in the Universe. A statistical analysis of FRBs shows 
that its rate should be equal to 10$^{-3}$ yr$^{-1}$  per a Milky Way type galaxy~\citep{Thornton}. At the same time,  the studies based on the population synthesis of binary stars usually yield substantially 
lower star merger frequencies: 10$^{-6}$--10$^{-4}$ yr$^{-1}$ per galaxy~\citep{Clark, Hils, PS, Bethe, Portegies,  Fryer,  Kalogera00, Kalogera01, Belczynski2,  Kim}.
The statistics of GRBs yields even lower occurrence frequencies. However, the latter estimate depends on the beaming angle of GRBs, which is very poorly defined \citep{Sari, Coward}.
At the same time, \citet{Totani} points out that theoretically estimated NS coalescence rates reported, in particular, by \citet{Thornton}, characterize the average or present-day 
occurrence frequencies, whereas the actual NS merger rate varied dramatically over the age of the Universe.

This study is based on the results of the Scenario Machine (see a monography; \citet{Lipunov96a}), which is the first and most mature population synthesis program aimed at simulating the evolution of a large number of single 
and binary stars. We focus on estimates of FRB rate in the framework of the NS+NS binary merger model. 
In Section 2, we give a brief review of the Scenario Machine population synthesis. In Section 3, we calculate the NS merger rate for more plausible star formation rate (SFR) and compare with observed frequency 
of FRB events. In Section 4, we discuss NS mergers as a result of the long binary evolution process and put some constraints on the natal kick velocities of NSs. In Section 5, we compare our estimates of NS+NS and 
BH+BH merger rate with the upper limits for these rates set by the lack of gravitational-wave event detection in LIGO experiment. In Section 6, we summarize our results.

\section{Scenario machine and neutron star merger rate}
The idea of population synthesis is to use our (not always complete) understanding of the evolution of binary stars, 
including relativistic stages, to generate millions computer-simulated artificial binaries in order to construct an artificial Universe and estimate the occurrence rates of various types of binaries and the 
associated cataclysmic events: supernova (SN) explosions, mergers, disruptions, etc. \citep{Kornilov83}.
It allows one to study the evolution of a large ensemble of binaries, to estimate the number of binaries at different evolutionary stages, and to explain the observational data.

The very first population synthesis computations performed using the Scenario Machine showed that the current NS merger rate in a galaxy like ours is equal to several events in 4000 years 
(see Fig.~1, case e~--~1 yr$^{-1}$, $<$ 20 Mpc in \citet{Lipunov87}). Subsequent computations demonstrated a strong evolution of merger rate with the age of the Universe 
\citep{Lipunov95} and kick velocity \citep{Lipunov97}.

Currently, about a dozen teams are implementing a Monte Carlo version of the population synthesis \citep{Abadie10}. We use the Scenario Machine population synthesis program (see a monography; \citet{Lipunov96a}). The Scenario Machine, first, incorporates not only the evolution of binary stars, 
but also the rotational evolution of compact stellar objects: NSs and white dwarfs (WDs). Furthermore, because of a certain number of obscure issues in our understanding of the 
evolution of binary stars: poor understanding of common-envelope stages, possible collapse anisotropy, poor knowledge of the initial component mass ratio distribution in binaries -- our 
approach uses the idea of optimization of all model parameters concerning NS evolution. For example, before computing the NS coalescence rate we optimize the parameters of evolution by 
comparing our artificial Galaxy with observations \citep{Lipunov96b}. We make sure that (1) the number of simulated binaries of various types agrees (approximately) 
with the corresponding number of actually observed binaries: binary radio pulsars, X-ray pulsars, black holes (BHs) with massive and low-mass companions, (2) the distribution of space 
velocities of  radio pulsars agrees with the corresponding distribution for actually observed radio pulsars, and (3) that the same is true for the number and parameters of radio 
pulsars with optical components, etc.

If the number of observed reference  quantities (the number of objects genetically related to NS+NS mergers) is sufficiently large, it is no longer important to what extent we might 
misestimate the hidden parameters such as the common-envelope parameter or the initial distribution of component mass ratios in binary stars. In other words, given a satisfactory 
explanation of the statistics of observed binary types we can be sure enough of the predicted statistics of unobserved events, such as NS mergers. 
Here we can formulate the 
following joke theorem (the theorem of population synthesis): the observed Universe of binary stars can be described to any desired degree of precision in terms of any, albeit incorrect, 
model with arbitrary large number of free parameters.

\begin{figure}
\includegraphics[width=84mm]{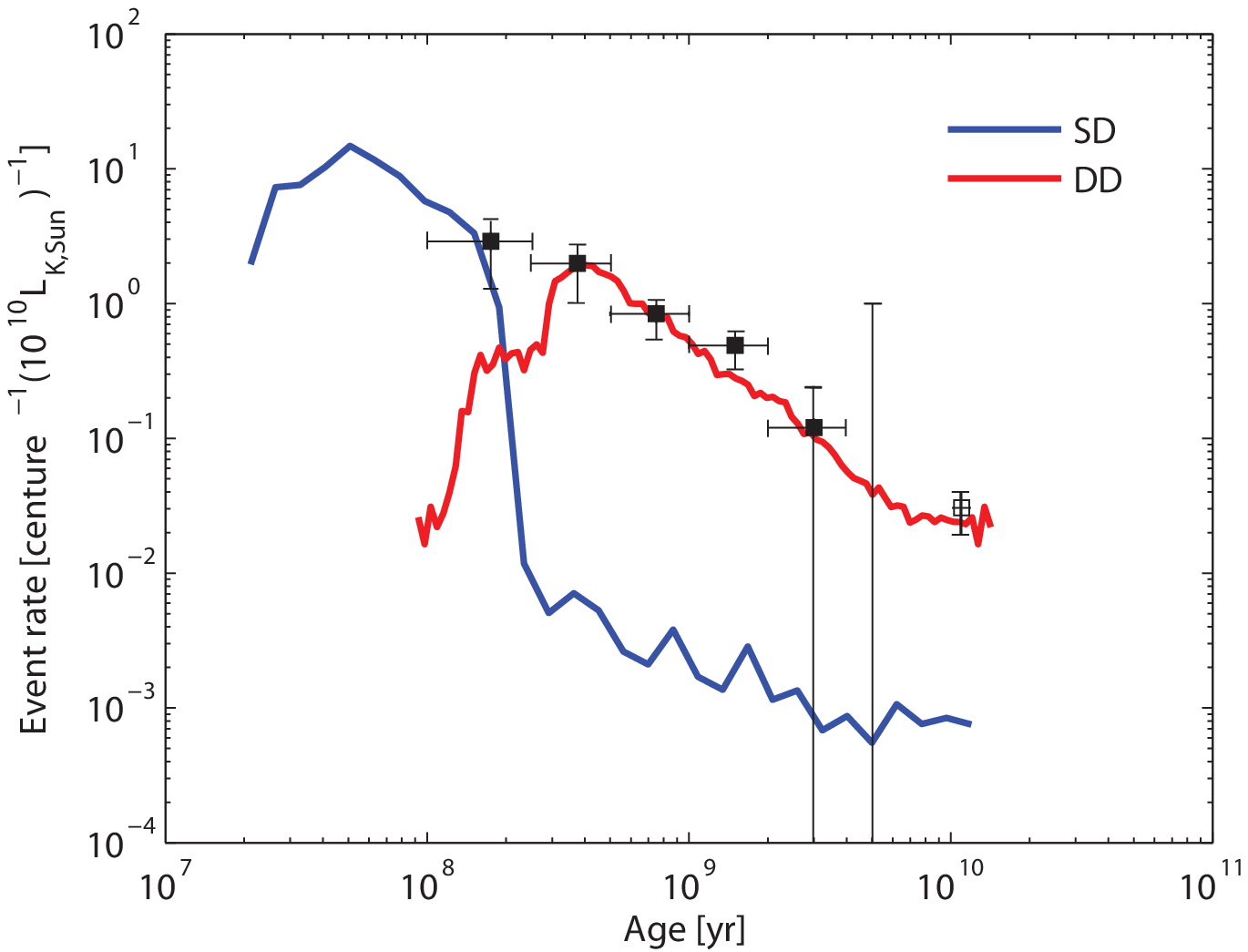}
 \caption{The SN Ia rate per century for a single starburst population whose total K-band luminosity is 10$^{10}$ L$_{K,\sun}$ at the age 
of 11 Gyr. The filled squares are the observational points. The open square is the observed SN Ia rate 
in elliptical galaxies in the local Universe \citep{Totani08,Mannucci}. The theoretical curves are the result of calculations \citep{Jorgensen} 
rescaled to the new units \citep{Lipunov11}. SD -- single degeneration, DD - double degeneration mechanisms.}
 \label{SN}
\end{figure}

The most striking confirmed predictions of the Scenario Machine include the discovery of radio pulsars paired with massive stars (predicted by \citet{Kornilov84, 
Lipunov94}; discovered by \citet{Johnston}), the discovery of strong evolution of the X-ray emission of galaxies \citep{Lipunov88,  Tatarintzeva, 
Lehmer} and its relation to the SFR~\citep{Lipunov96c}, prediction of the rate and nature of the evolution of mergers of WDs with 
the total mass above the Chandrasekhar limit (see Fig.~\ref{SN}; \citet{Totani08, Lipunov11}), and the discovery of dark energy signature in the statistics of GRBs 
(see Fig. 2; \citet{Lipunov95}).  The Scenario Machine continues to develope and one of the latest version is published in \citet{Lipunov2009}.

\begin{figure}
\includegraphics[width=70mm]{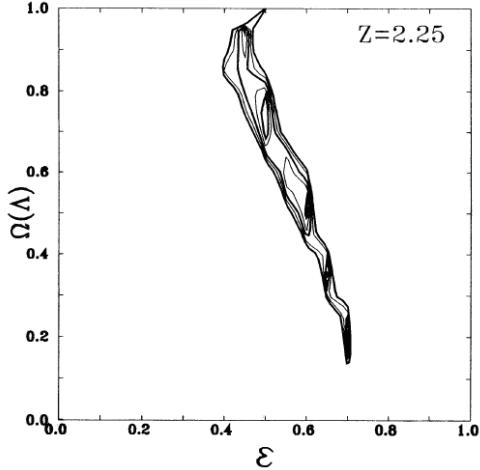}
 \caption{Dependence of dark-energy density on the fractional part of the luminous baryonic matter entering into the first-generation metal-poor stars ($\epsilon$) estimated assuming that star formation rate peaked at 
about z = 2.25. According to modern data, the peak is located at about $\sim$ 2--2.5. 
Current value of $\Omega_{\Lambda}$ = 0.7 \citep{Perlmutter} gives $\epsilon$ $\simeq$ 0.5, which seems reasonable.
The (90\%, 91\%, etc.) confidence level contours are shown for the $\omega^2$ 
test in the $\epsilon-\Omega_{\lambda}$ plane at z$_*$ = 2.25. A flat cosmological model and GRB spectral index s = 1.5 with the source evolution as in Fig.~\ref{NS} are 
adopted \citep{Lipunov95}.}
 \label{omega}
\end{figure}

\begin{figure}
\includegraphics[width=84mm]{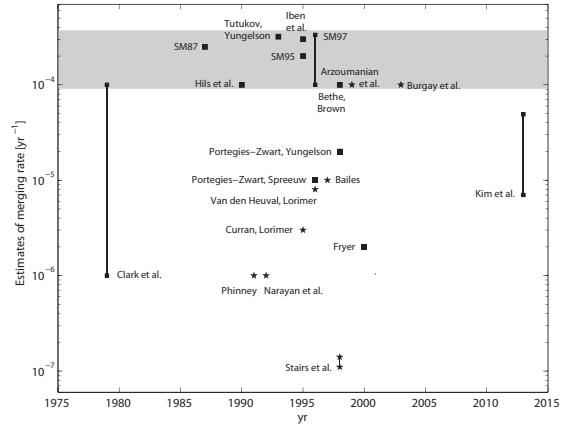}
 \caption{Estimated merger rates for NS at the present epoch normalized per year per galaxy with a constant SFR of 1M$_{\sun}$ per year (like the Milky Way). 
The squares show the coalescence rate estimates obtained using the method of population synthesis (they are sometimes referred to as theoretical estimates). The asterisks show the ``observed'' estimates based on the 
statistics of binary radio pulsars. The gray region outlines the Scenario Machine predictions. Clark et al. -- \citet{Clark}; SM87 -- \citet{Lipunov87}; 
Hils et al. -- \citet{Hils}; Phinney -- \citet{Phinney}; Narayan et al. -- \citet{Narayan}; Tutukov, Yungelson -- \citet{Tutukov}; Iben et al. -- \citet{ITY}; SM95 -- \citet{Lipunov95}; Curran, Lorimer -- \citet{Curran}; 
SM97 -- \citet{Lipunov97}; Portegies-Zwart, Spreeuw -- \citet{PS}; Van den Heuvel, Lorimer -- \citet{Heuvel}; Bailes -- \citet{Bailes}; Portegies-Zwart, Yungelson -- \citet{Portegies}; Bethe, Brown -- \citet{Bethe}; 
Stairs et al. -- \citet{Stairs}; Arzoumanian et al. -- \citet{Arzoumanian}; Fryer -- \citet{Fryer}; Burgay et al. -- \citet{Burgay}; Kim et al. -- \citet{Kim}.}
\label{rate}
\end{figure}

Figure \ref{rate} briefly outlines the history of NS coalescence rate normalized per Milky Way like galaxy and referred to the present epoch. We do not include in Fig. \ref{rate} studies 
covering a range of values of $\sim$2--3 orders of magnitude \citep{Kalogera01, Kalogera04, Belczynski2} except the very first theoretical consideration performed by \citet{Clark}.
As it is evident from the figure, the NS coalescence rate provided by the Scenario Machine -- 1/3000--1/10000 yr$^{-1}$ per galaxy -- has 
remained unchanged since 1987, whereas observational data, on the contrary, gradually approach the ``theoretical'' estimate of the Scenario Machine. 
In the paper \citet{Bethe}, the predicted rates for NS+NS and BH+NS binaries are 10$^{-5}$ and 10$^{-4}$ yr$^{-1}$ per galaxy, respectively. However, we would like to stress that \citet{Bethe} considered BHs 
with mass close to the Oppenheimer--Volkoff limit ($\sim$2--2.5~M$_{\sun}$). Such low-mass BHs are formed due to hyperaccretion during common-envelope stage \citep{Chevalier}. 
In case of Fig.~\ref{rate} low-mass BH+NS mergers are equivalent to NS+NS mergers because the horizon distance for these two types of binaries is approximately the same. 
Consequently, 10$^{-4}$ yr$^{-1}$ per galaxy is a real estimate for NS+NS mergers.

The LIGO experiment will detect nearby ($<$100~Mpc) coalescence 
events  and therefore we are primarily interested in the estimate of the current coalescence rate. To estimate the total number of events to be recorded in LIGO and other ground-based 
experiments, we must sum up the star-formation rate within the volume of the gravitational-wave horizon and add a small contribution provided by elliptical galaxies.

Note that population synthesis depends to an important hidden parameter, the so-called kick velocity. Here, we mean the natal kick velocity, in other words a velocity that NS receives due to anisotropy of SN explosions. However, this parameter can be constrained substantially by comparing the results 
of population synthesis with the statistics of NS+NS and NS+WD binaries. This is understandable. If the collapse of an SN is highly anisotropic then the probability for 
the binary to be preserved after the explosion is very low. The collapse anisotropy is especially critical for the number of NS binaries (i.e., Taylor type radio pulsars), 
because such systems form as a result of two supernova explosions in the same binary. A special analysis performed in 1997 showed that the collapse anisotropy cannot be very high: 
100--150 km s$^{-1}$ (Fig.~\ref{kick_matlab};~\citet{Lipunov96b}, see also fig.~1 in \citet{Lipunov97}). With such a kick velocity the current coalescence rate should be equal to 1/3000--1/5000 yr$^{-1}$ per $10^{11}$ M$_{\sun}$ in a Milky Way type galaxy.
However, this is the coalescence rate at the present epoch. As \citet{Totani} pointed out, the coalescence rate could have been significantly higher in the past, and we appear to be 
dealing with redshifts about 0.5--0.1 if FRBs are cosmological events. It was first shown by the computations performed using the Scenario Machine that the coalescence rate evolves 
dramatically even if normalized to the rest frame \citep{Lipunov95}.

\begin{figure}
\includegraphics[width=84mm]{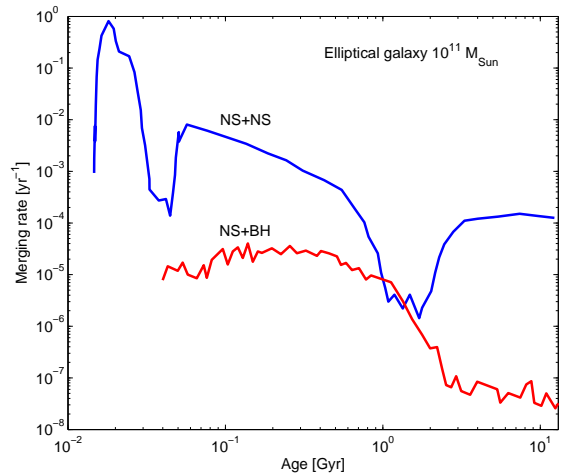}
 \caption{Temporal evolution of NS+NS (blue curve) and NS+BH (red curve) coalescence rates calculated for 2$\times$10$^8$ binaries and normalized to a model elliptical galaxy with baryonic mass $10^{11}$ M$_{\sun}$ \citep{Lipunov95}. 
The gaps in the NS+NS merging rate are appeared because in initial binary systems one distinguishes different cases of mass transfer depending on the nuclear evolutionary state of the star \citep{Webbink}.}
 \label{NS}
\end{figure}

\begin{figure}
\includegraphics[width=81mm]{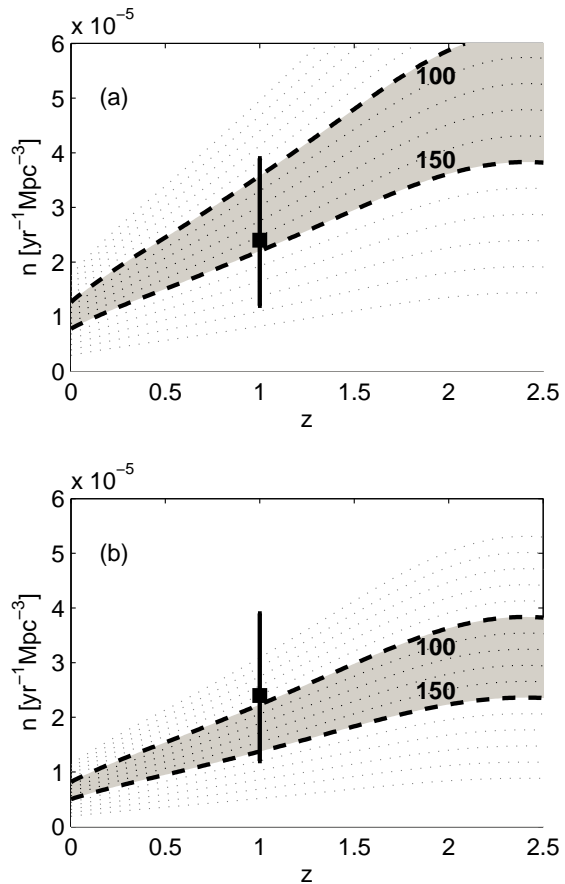}
 \caption{Estimated merger rates for NS normalized per year per comoving Mpc$^3$ for SFR based on the adopted \citet{Salpeter} (a) and \citet{Baldry} (b) IMF 
for kick velocities in the 100--150 km s$^{-1}$ interval. 
The black square shows the estimated FRB rate interpreted as a burst rate per unit comoving volume out to z = 1.}
 \label{n}
\end{figure}

\begin{figure}
\includegraphics[width=84mm]{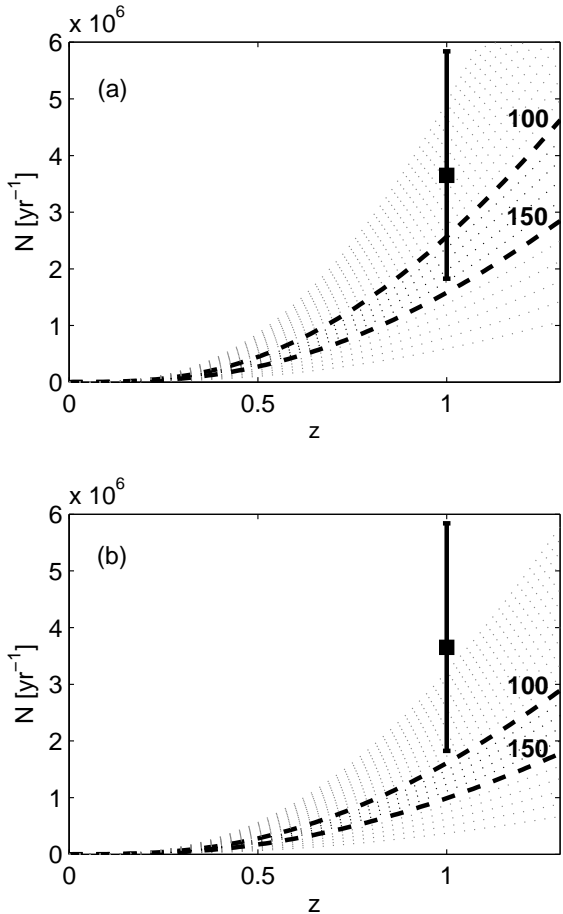}
 \caption{The number of NS+NS mergers per year inside the sphere of redshift z for SFR based on 
the adopted \citet{Salpeter} (a) and  \citet{Baldry} (b) IMF for kick velocities in the 100--150 km s$^{-1}$ interval. The black square shows the estimated FRB rate \citep{Thornton}.}
 \label{N}
\end{figure}

Based on these computations (Fig.~\ref{NS}) and the modern data on SFR we derived the dependence of the NS coalescence rate in the Universe. We used the compilation of $UV$, 
$FIR$, radio and H$\alpha$ SFR data from \citet{Hopkins} that were fitted with parametric form of \citet{Cole}: $\dot{\rho_{\star}} = (a + bz)h/[1+(z/d)^c]$, where $h = 0.7$ is Hubble constant. 
SFR was determined up to z $\sim$ 6. However, we extended it up to z = 10, because the SFR measurements at redshifts 6$<$z$<$10 \citep{Bouwens04, Bouwens05, Bouwens06} showed 
that the decline in SFR seen near z~=~6 continues to higher redshifts. 
We used two sets of parameters for two possible extreme initial mass function (IMF): the modified Salpeter IMF with the high-mass power-law slope of -1.35 \citep{Salpeter}
and the IMF of \citet{Baldry} with the high-mass power law slope of -1.15, because other IMF approximations lie between this two \citep{Hopkins}.
We then computed the NS coalescence rate, $n$, per comoving Mpc$^3$ at time z for this type 
of SFR and the cumulative NS merger rate, $N$, for different kick velocities obtained from the Scenario Machine simulations (see Figs~\ref{n},~\ref{N}) and find that 
\begin{equation}
n(t) = \int_{0}^{t} SFR(\tau)G(t-\tau)d\tau;~t \to z.
\end{equation}
Here, $G(t)$ is the NS merger rate in a sample galaxy after an instantaneous star formation at the moment $t$ = 0.
The number of events per unit time within the sphere of redshift z is
\begin{equation}
N(z) = 4\pi\int_{0}^{z} \frac{n[t(z)]D(z)^2dD}{1+z},
\end{equation}
where $D(z)$ is a comoving distance.

\section{Neutron star mergers and FRB}
The most important remaining issue is to explain the observed occurrence frequency of FRB events. \citet{Thornton} point out that the occurrence rate of radio bursts is 
substantially higher than the predicted rate for NS mergers. However, Totani et al. 2013 note that the discrepancy practically disappears if the increase of the NS 
coalescence rate in the past due to higher SFR in the Universe is taken into account. Actually, NSs should have sufficiently strong magnetic fields for the 
orbital triggering mechanism \citep{Lipunov96} to operate. Unfortunately, the magnetic field dissipation time-scale for an NS is poorly known. Radio pulsar studies 
appear to imply that magnetic field dissipates during the first 10 Myr after the birth of the NS. However, the decay of radio pulsars may be due to the change of the angle 
between the magnetic and rotation axes rather than the dissipation of magnetic field. On the other hand, accreting NSs in binaries -- X-ray pulsars -- show that, judging by 
the companion, the magnetic field has an age and strength of $10^8$ years and $10^{12}$ G, respectively \citep{Lipunov92}.
The computations of the coalescence rate evolution as a function of the age after star formation show that 70 per cent of coalescing NSs have ages 
below 100 Myr \citep{Lipunov95}.

Let us now compare the resulting NS coalescence rate using the Scenario Machine to the FRB rate. 
The FRB rate should be $1.0^{+0.6}_{-0.5}\times10^4$ day$^{-1}$ sky$^{-1}$, where the 1-$\sigma$ uncertainty assumes Poissonian statistics \citep{Thornton}.
If it is interpreted as a burst rate per unit comoving volume out to z = 1, then for FRB rate we find $2.4^{+1.5}_{-1.2}\times10^{-5}$ yr$^{-1}$ per Mpc$^{3}$.
This result is consistent with the Scenario Machine estimate of NS+NS mergers (Fig.~\ref{n}).
The cumulative number of NS+NS mergers per year inside the sphere of redshift z and FRB rate are presented on Fig.~\ref{N}. 
The observed FRB rate is in better agreement with the model using SFR based on the adopted \citet{Salpeter} IMF, than the \citet{Baldry} IMF based one.

The concrete mechanism of the burst could be the revival of pulsar mechanism during the last revolutions before the coalescence, when the orbital frequency reaches several kHz 
\citep{Lipunov96}. The burst duration agrees well with the observed values. As for the observed fluxes, we could write down many formulas. However, the development of the 
theory of radio pulsars shows that it makes no sense to use any formulas for estimating radio emission, and that empirical data combined with a simple model of multidipole losses 
yield much more reliable results. A more correct approach should involve the use of the quadrupole formula \citep{Lipunov96}, however, the losses it yields at maximum agree 
with those computed in terms of the dipole model. It is therefore quite safe to use the dipole formula, i.e., assume that the rate of emission is proportional to the fourth power of 
frequency. Given that the Crab nebula with its 30 Hz rotation frequency emits a radio flux of 1000 Jy,  a similar Crab with a 1 kHz frequency should produce a $10^6$ greater power 
output. This means that such an object would produce a 1 Jy flux from a distance of 70 Mpc. To explain the observed intensity of FRB, a 1~Jy flux should be observed from a source 
at 2 Gpc, i.e., 30 times more distant. To this end, we must assume that the FRB magnetic field should be stronger by the same factor of 30. An analysis of the distribution functions 
of radio and X-ray pulsars shows that up to one half of all NSs may possess such magnetic fields \citep{Lipunov92}. A factor of 2 is the approximate accuracy of the estimated 
FRB occurrence rate.   

\section{Neutron star mergers\\ and Short GRB}
Recently, \citet{Abadie10} published a paper about the predicted observed relativistic binary coalescence rates detectable by gravitational wave detectors. 
Most of the paper is dedicated to predicting the detection rates for future (advanced) LIGO versions based on observational data and theoretical predictions.
The wide range of the resulting NS merger rates is caused by the large uncertainties of kick velocities assumed in the theoretical models \citep{Kalogera04, Belczynski08} which are being used by the above authors,
although previously it was shown that high kick velocity values contradict to observable fraction of NS+NS binaries among the total number of pulsars on the sky (see fig.~1 in \citet{Lipunov97}).
For example, the used models completely 
lack the first and most complete, incorporating all the observed stages of the evolution of binaries with relativistic components and of ordinary stars, computations of the NS 
coalescence rates performed long before the studies used in that paper were published (for example, see a monography;~\citet{Lipunov96a}).

We reiterate that NS mergers are a result of the long binary evolution process starting from two optical main-sequence stars, through the two SN explosions, and involving ejector, 
propeller, accreting evolution of the NSs, and after the pure evolution of the two NSs controlled the angular dissipation of gravitational waves. Before predicting the NS and 
BH coalescence rate one has to: (1) explain the existence and statistical properties (the number and characteristic luminosity) of X-ray pulsars with massive OB-type stars, 
(2) the statistics of propellers and ejectors in massive systems, (3) the statistics of BHs in massive binaries, (4) the existence of binary radio pulsars with NSs and 
WDs, and (5) the lack of such systems with BHs. The answer to the question of the NS coalescence rate actually depends on what happens in low-mass stars. 
The point is that one of the most obscure issues in the evolution of binary stars -- the common-envelope stage -- shows up most conspicuously during the formation of low-mass cataclysmic 
variables, which is  often controlled by gravitational waves.

\begin{figure}
\includegraphics[width=90mm]{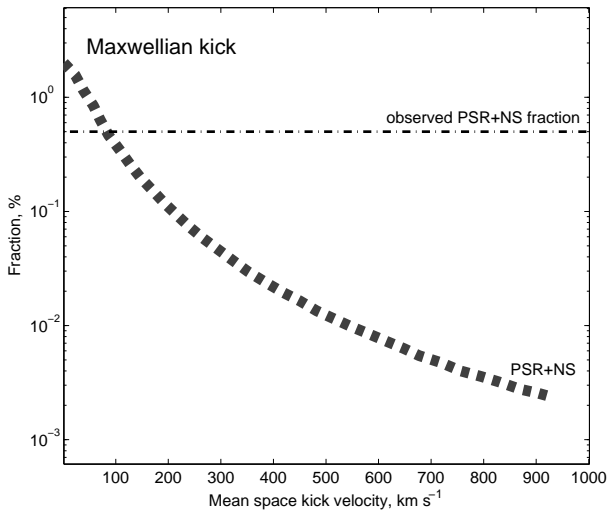} 
\caption{Binary NS+NS fraction among the total number of pulsars (PSR) as a function of the mean kick velocity for a Maxwellian distribution. The horizontal dot dashed line shows the observed fraction of 
NS+NS binaries among the total number of pulsars \citep{Lorimer08, Lorimer12, Ferdman}. If the natal kick velocity is significantly more than 150 km s$^{-1}$ you cannot explain the observed fraction of NS+NS mergers.}
 \label{kick_matlab}
\end{figure}

\begin{figure}
\includegraphics[width=100mm]{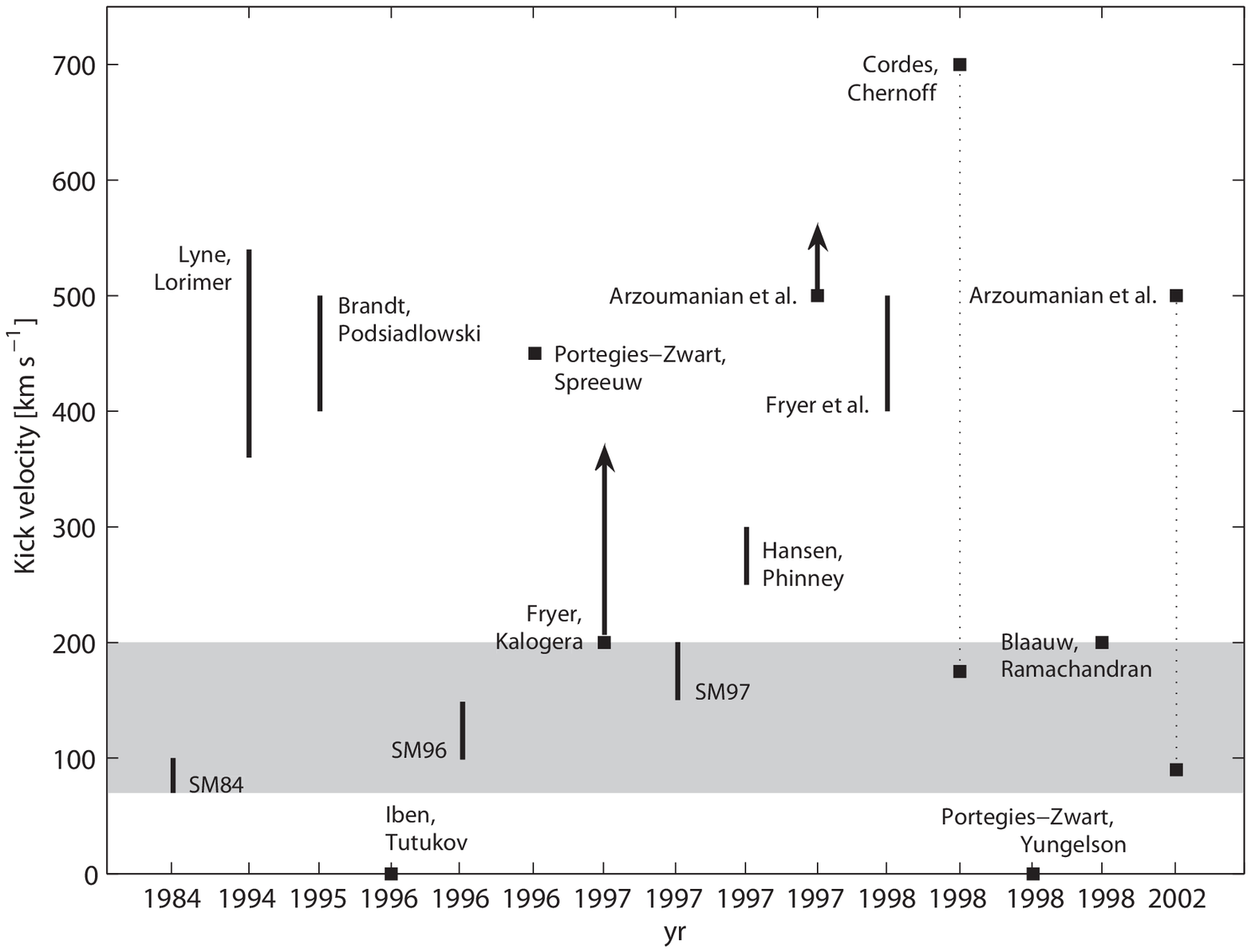} 
\caption{Kick velocity estimation by different authors. The gray region outlines the Scenario Machine predictions. SM84 (for $\delta$-function distribution) -- \citet{Kornilov84}; Lyne, Lorimer -- \citet{Lyne}; Brandt, Podsiadlowski -- \citet{Brandt}; Iben, Tutukov -- \citet{IT}; 
SM96 -- \citet{Lipunov96b}; Portegies-Zwart, Spreeuw -- \citet{PS}; SM97 -- \citet{Lipunov97}; Fryer, Kalogera -- \citet{Fryer}; Hansen, Phinney -- \citet{HP}; Arzoumanian et al. -- \citet{Ar}; 
Fryer et al. -- \citet{FBB}; Cordes, Chernoff (two-component velocity distribution) -- \citet{Cordes1998}; Portegies-Zwart, Yungelson -- \citet{Portegies}; Blaauw, Ramachandran -- \citet{BR}; Arzoumanian et al. 
(two-component velocity distribution) -- \citet{ACC}.}
 \label{kik}
\end{figure}

The first computations of the NS coalescence rate made using the Scenario Machine date back to 1987 \citep{Lipunov87} and the results have remained 
practically unchanged until now. For example, after determining the optimum evolution parameters \citep{Lipunov96b}, we performed a detailed analysis of the 
effect of collapse anisotropy and confirmed our earlier conclusions that the kick velocity cannot exceed 100--150 km s$^{-1}$ for a Maxwellian distribution (\citet{Lipunov96b}, see also fig.~1 in \citet{Lipunov97}). 
Otherwise, you cannot explain the observed NS+NS fractions among the total number of pulsars (Fig.~\ref{kick_matlab}).
It must be emphasized that the characteristic velocity depends on the form of the natal kick distribution. Thus earlier a number of authors \citep{Lyne, ACC} proposed 
non-Maxwellian distributions with a higher fraction of higher kick velocities (Fig.~\ref{kik}). This is especially important for explaining the observed distribution of radio pulsar velocities. 
However, the particular form of the distribution is of little importance for this study, because in all the distributions considered the number of small natal kicks is approximately 
the same to within a factor of 2.

\section{LIGO silence}
\begin{figure}
\includegraphics[width=84mm]{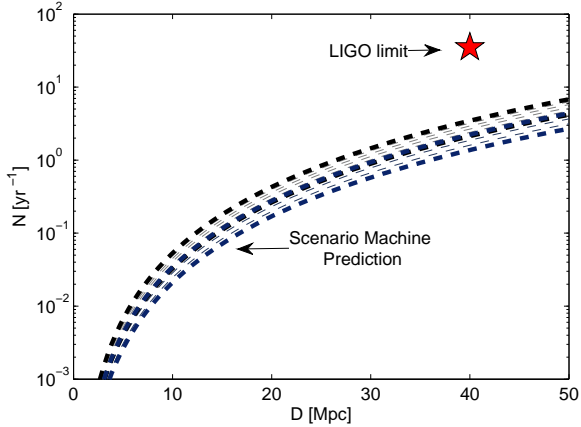}
 \caption{The number of NS+NS mergers per year inside the sphere of distance $D$ in terms of the Scenario Machine prediction for kick velocities in the 100--150 km s$^{-1}$ interval. 
The black and blue curves correspond to modified \citet{Salpeter} and \citet{Baldry} IMF, respectively.
The asterisk shows the LIGO S6 limit \citep{Abadie12}.}
 \label{ligo}
\end{figure}

If we indeed plan to associate FRB with the coalescence of NSs, we must adopt our average estimate of the NS coalescence rate at distances out to 40 Mpc 
(the horizon distance for LIGO S6; \cite{Abadie12}), i.e. $10^{-5}$ yr$^{-1}$ per Mpc$^3$. Given the small volume of this region, we obtain a rate of two events per year, which is 
consistent with non-detection of gravitational pulses in the LIGO experiment (see Fig.~\ref{ligo}). The upper limit for NS coalescence rate set by the lack of gravitational-wave 
event detection in LIGO experiment is $1.3\times10^{-4}$ yr$^{-1}$ per Mpc$^3$ (35 events per year; \cite{Abadie12}).
LIGO silence is therefore consistent with astronomical observational data (Fig.~\ref{ligo}).
However, since a typical BH is formed with a mass higher than the NS mass and the detection volume is proportional to $M^{5/2}$, where $M$ is a ``chirp'' mass of the binary system, 
the expected detection rate of binary BH by LIGO is 10--100 times higher than the binary NS merging rate \citep{LipunovNA}. According to this for BH mergers we obtain more than 20 events per year that slightly contradict LIGO limit 
for such type of events (20 events per year; \cite{Abadie12}). 

The optimum horizon distance for NS+NS events in Advanced LIGO is assumed to be equal 445 Mpc \citep{Abadie10}. If LIGO reaches such sensitivity, the gravitational waves from 
merging NS are surely be detected. Within such distances the Scenario Machine predicts several thousand events per year. The cumulative rate of events per year in the volume within 
500 Mpc can be approximately described by formula: $(4\pm2)\times10^{-5}$ ($R$/Mpc)$^3$ yr$^{-1}$, where $R$ is the horizon distance in Mpc.

\section{Discussion}
We present for the first time the evolution of NS coalescence rate as a function of redshift in terms of a reasonable star formation function in the Universe. For a kick 
velocity of 100--150 km s$^{-1}$ this function yields an average coalescence rate of 1/500--1/2000 yr$^{-1}$ per galaxy in the comoving volume corresponding to z = 0.5--1, which is consistent 
with observational estimates \citep{Thornton}. Our analysis  is based on the results of population-synthesis studies performed in 1995-1997 using the Scenario Machine. We accept 
the criticism from the teams that take into account the following five effects in their computations:  mass exchange in binaries, common-envelope stages, angular momentum 
carry-over by matter, magnetic stellar wind, and gravitational waves, anisotropy of SN explosions (the kick velocity effect), and rotational evolution of magnetized 
NSs and WDs. In addition, these computations should correctly describe the statistics of 10 types of binaries or their observed properties: (1) binary radio 
pulsars (with NSs, WDs, and optical companions), (2) X-ray binaries with massive OB-type stars, (3) evolution of the Type Ia SN rate as a function of the age 
of the Universe, (4) distribution of intermediate polars, (5) velocities of single radio pulsars, (6) total X-ray luminosity of galaxies, (7) BHs paired with massive 
companions (Cyg X-1), and (8) statistical properties of millisecond pulsars spun up by accretion. 

The fact that we did not detect any gravitational waves from NS mergers in LIGO search is consistent with our astronomical predictions but BH mergers could already be registered.

We finally argue that there are no discrepancies between the NS coalescence rate and FRB occurrence rate. 
Furthermore, we predict a certain (about 20 per cent) anisotropy of FRB (directivity pattern).

\section*{Acknowledgments}
One of the authors (VL) is grateful to late Leonid Grishchuk for his initiating advice to continue publications on the population synthesis of stars. I want to point out that is was 
Leonid Petrovich Grishchuk (together with Kip Thorne) who prompted us to study the NS coalescence rate in the 1980s. We acknowledge I.E. Panchenko for comments and discussion.
We thank the Extreme Universe Laboratory of Lomonosov Moscow State University Skobeltsyn Institute of Nuclear Physics. This work was partially supported by funds from Megagranta no. 11.634.31.0076. 
This work was supported in part  by M.V. Lomonosov Moscow State University Program of 
Development, by Ministry of Education and Science of the Russian Federation (agreement 8415 on 27 August 2012), and the ``Dynasty'' Foundation of non-commercial programmes.

\label{lastpage}

\end{document}